\documentclass[10pt,conference]{IEEEtran}

\usepackage{amsmath,amssymb,bm}
\usepackage{booktabs}
\usepackage{array}
\usepackage{multirow}
\usepackage{graphicx}
\usepackage{cite}
\usepackage{microtype}
\usepackage{url}
\graphicspath{{.}{/mnt/data/}}

\title{Objective Diversification of Dynamic Hedging}
\author{
\IEEEauthorblockN{Brian Sing Fan Chan}
\IEEEauthorblockA{School of Mathematics and Statistics\\
The University of Sydney\\
Sydney, Australia}
\and
\IEEEauthorblockN{Carlin Chu}
\IEEEauthorblockA{Department of Computer Science\\
The Hang Seng University of Hong Kong\\
Hong Kong}
}

\begin{document}
\maketitle

\begin{abstract}
A hedging objective reflects the risk preference of the hedger, so the conventional approach is to estimate the hedge ratio directly under the chosen objective. This paper asks whether that estimate should be used alone. We construct five objective-specific rolling hedge-ratio estimators---minimum variance, least absolute deviation (LAD), downside semivariance, lower-tail quantile, and Expected Shortfall (ES)---from the same historical stock--market returns, and combine the resulting estimates under a fixed final evaluation objective. Separate multi-output ExtraTrees models predict the within-stock-date ranks of the five candidates under MSE, LAD, downside semivariance, and 10\% quantile loss. The design uses pre-specified 60-, 90-, 120-, and 180-day windows with objective-matched rolling estimators as benchmarks. Adaptive weighting improves the corresponding loss in all sixteen objective--window comparisons. Equal weighting, which uses the same objective-specific hedge ratios without state variables or learned reweighting, already improves downside and quantile hedging. The results therefore suggest that estimates generated under other objectives can be useful even when the final hedging preference and evaluation objective are fixed, while adaptive weighting can provide a further improvement.
\end{abstract}

\section{Introduction}

A standard market hedge offsets part of an individual stock's systematic return with a broad market instrument. Let $r_{i,t+1}$ denote the return of stock $i$ on target date $t+1$ and $r_{m,t+1}$ the return of the market hedge instrument. A hedge ratio $\beta_{i,t+1}$ chosen using information available through formation date $t$ produces the realized hedged return
\begin{equation}
    e_{i,t+1}=r_{i,t+1}-\beta_{i,t+1}r_{m,t+1}.
    \label{eq:hedged_return}
\end{equation}
Under the conventional minimum-variance criterion, the population hedge ratio is
\begin{equation}
    \beta_{i,t+1}^{\mathrm{MV}}
    =
    \frac{\operatorname{Cov}_{t}(r_{i,t+1},r_{m,t+1})}
    {\operatorname{Var}_{t}(r_{m,t+1})}.
    \label{eq:minvar_beta}
\end{equation}
Rolling OLS is therefore a natural dynamic benchmark. More generally, dynamic hedging has been studied through time-varying covariance models and other conditional estimators \cite{kronersultan1993,engle2002}, while recent machine-learning work estimates time-varying hedge ratios directly from historical stock--market information and evaluates them through realized hedging performance rather than against an unobservable ``true'' beta \cite{liu2024neuralbeta}.

This paper focuses on a different source of flexibility: the \emph{hedging objective}. The objective reflects the aspect of risk that the hedger wishes to control. Once an objective is chosen, the natural approach is to estimate the hedge ratio directly under that objective. Minimum variance, LAD, downside, quantile, and ES criteria, however, can generate different scalar hedge-ratio estimates from exactly the same return history. We therefore ask whether, for a fixed final hedging objective, it is sufficient to rely only on the estimate obtained directly under that objective, or whether estimates generated under other objectives can help construct a better hedge for the same criterion.

We study this question through \emph{objective diversification}. For each stock-date and estimation window $W$, five rolling estimators use the same historical observations and hedge specification but different estimation objectives to generate five candidate hedge ratios. The diversity therefore comes from the objective function rather than from changing the model class. A second-stage combination model then learns how to combine these estimates under a given evaluation objective. The machine-learning layer does not estimate an unconstrained hedge ratio directly; it determines the weights assigned to the objective-specific candidates. The complete procedure is summarized in Fig.~\ref{fig:framework}.

The second stage is estimated separately for four loss criteria: MSE, LAD, downside semivariance, and 10\% quantile loss. During training, the five candidate hedge ratios are applied to the realized target-date returns, their one-day hedging losses are evaluated under the chosen criterion, and those five losses are ranked within each stock--date. This five-dimensional rank vector is the supervised target of the multi-output ExtraTrees model. Using only information available at the formation date, the model predicts the next-period ranks, which are then converted into smooth convex weights. Equal weighting and hard selection provide complementary controls for interpreting the adaptive results. In particular, Equal-weight combines the same objective-specific hedge ratios without using the broader formation-date state variables, helping distinguish the value of the candidate representation from the additional value associated with state-dependent weighting.

The empirical design uses four pre-specified rolling estimation windows,
\begin{equation}
    W\in\{60,90,120,180\}\ \text{trading days}.
    \label{eq:windows}
\end{equation}
These horizons fall within commonly used short-to-medium rolling windows in empirical finance for hedge-ratio, beta, and related market-exposure estimation \cite{brailsford2001,jermann2022,cillo2018,wan2021}.
Within each combination-objective panel, Adaptive-weight, Hard selection, and Equal-weight are compared with a corresponding single-objective rolling hedge using the same estimation window. Squared-loss performance is benchmarked against the conventional rolling minimum-variance (OLS) hedge, while LAD, downside semivariance, and quantile loss are compared with their respective rolling estimators. This construction keeps the estimation window and evaluation criterion fixed when assessing the incremental value of combining heterogeneous candidate hedge ratios.

Across the 60-, 90-, 120-, and 180-day rows, Adaptive-weight has positive Obj. Gain in all sixteen objective--window comparisons. The effect is strongest under downside and Q10 criteria and smaller under MSE and LAD. Equal weighting is already highly competitive under downside and Q10, while it is weaker under MSE and mixed under LAD. These results indicate that the hedge ratio estimated directly under the final objective need not exhaust the value available from the same return history: estimates generated under other objectives can also contribute to the final hedge. The cumulative-loss figure shows positive endpoint savings across all four objective panels and all four estimation windows.

The paper makes three contributions. First, it studies objective diversification as a source of hedge-estimator diversity by varying the estimation objective while holding the return information and hedge specification fixed. Second, it tests whether objective-specific estimates can improve hedging under a fixed final criterion relative to estimating the hedge ratio directly under that same criterion. Third, it uses equal weighting and state-dependent learning to separate the value of combining objective-specific estimates from the additional value of adaptive weighting.

\section{Related Work and Positioning}

The classical hedging literature evaluates hedge ratios by the reduction in realized risk of the hedged position, with variance hedge effectiveness providing a standard economic measure \cite{ederington1979}. Reviews of the optimal-hedge-ratio literature emphasize that the hedge ratio depends on the risk objective and document minimum-variance and semivariance-based formulations as established alternatives \cite{chen2003review}. Dynamic hedge-ratio estimation extends this logic by allowing the conditional covariance between an asset and its hedge instrument to change through time \cite{kronersultan1993,engle2002}. Rolling OLS remains a transparent benchmark because it updates the market beta from recent observations while retaining the conventional minimum-variance objective.

Alternative objectives also have direct precedents in financial risk management and hedging. Absolute-deviation criteria have long been used as robust alternatives to least squares for market characteristic-line estimation \cite{sharpe1971} and have been applied directly to hedging-portfolio selection \cite{liu2009mad}. Downside-risk hedge ratios based on semivariance or lower partial moments have been studied explicitly in futures hedging \cite{lientse2000,chen2001gsv}. Quantile regression has been used to construct quantile-specific optimal hedge ratios \cite{lien2016quantile}, while minimum-CVaR/Expected-Shortfall hedges have been considered as tail-risk alternatives to minimum-variance hedging \cite{harris2006var}.

A separate literature asks whether nonlinear models can estimate dynamic market exposure more flexibly. Recent deep-learning work such as NeuralBeta \cite{liu2024neuralbeta} treats the hedge ratio as the output of a learned mapping from historical stock--market data. Our setting differs because the component estimators remain simple rolling estimators and machine learning is reserved for the combination problem.

The methodological motivation is also related to objective-diverse forecasting. Different objective functions can produce different predictions even when the inputs and model class are otherwise similar, and combining those predictions can be useful \cite{dahal2026}. In dynamic hedging, this creates a distinct source of estimator diversity: the same return information is mapped into several hedge-ratio estimates by changing the risk objective. Our approach therefore lies between relying on a single objective-specific hedge-ratio estimator and predicting the hedge ratio directly with machine learning.
\section{Objective-Diversified Adaptive Hedging}

\begin{figure*}[!t]
    \centering
    \includegraphics[width=1.2\columnwidth]{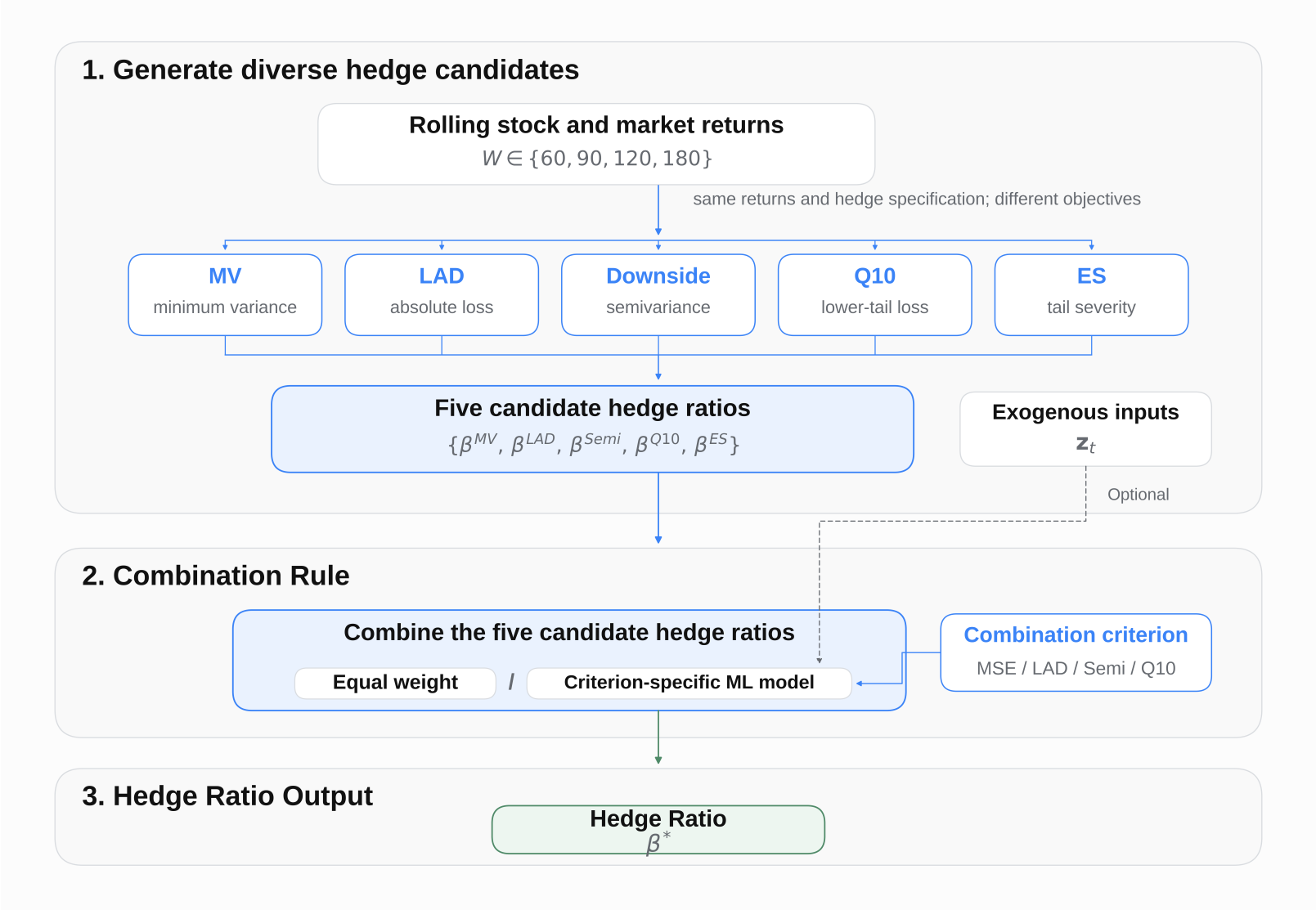}
    \caption{Objective-diversified dynamic hedging framework. Five objective-specific hedge ratios are estimated from a common rolling information set and combined using either equal weighting or a criterion-specific learning rule.}
    \label{fig:framework}
\end{figure*}

Figure~\ref{fig:framework} summarizes the proposed framework. The method separates the objectives used to generate candidate estimates from the criterion used to evaluate the final hedge. The same rolling stock--market return history is mapped into five hedge-ratio candidates under different hedging objectives. These candidates are then either equally weighted or combined using a criterion-specific learning model that incorporates formation-date state information. The resulting combination produces the final hedge ratio.

\subsection{Timing and rolling information sets}

For each stock $i$, the hedge applied on target date $t+1$ is formed using information available no later than formation date $t$. For a chosen estimation window $W$, let $\mathcal W_{i,t}^{(W)}$ contain the most recent $W$ valid paired stock and SPY returns. A component hedge ratio is formed only when the required history is available; missing returns are not forward-filled.

For a candidate hedge ratio $h$, define the in-window hedging residual
\begin{equation}
    e_{i,s}^{(W)}(\beta)=r_{i,s}-\beta r_{m,s},
    \qquad s\in\mathcal W_{i,t}^{(W)}.
    \label{eq:residual}
\end{equation}
\subsection{Objective-specific rolling hedge-ratio estimators}

Five hedging objectives generate five component hedge ratios. The first is the conventional rolling OLS minimum-variance hedge \cite{ederington1979,chen2003review},
\begin{equation}
\begin{aligned}
    \beta_{i,t+1}^{\mathrm{MV},W}
    &=
    \arg\min_{\beta}
    \widehat{\operatorname{Var}}_{s\in\mathcal W_{i,t}^{(W)}}
    \!\left(e_{i,s}^{(W)}(\beta)\right) \\[2pt]
    &=
    \frac{
    \sum_{s\in\mathcal W_{i,t}^{(W)}}
    (r_{i,s}-\bar r_{i,t})(r_{m,s}-\bar r_{m,t})
    }{
    \sum_{s\in\mathcal W_{i,t}^{(W)}}
    (r_{m,s}-\bar r_{m,t})^2
    }.
\end{aligned}
    \label{eq:rolling_ols}
\end{equation}
This is the familiar covariance-to-variance hedge ratio. The remaining components apply alternative loss and risk criteria to the same hedged-return series and the same rolling information set.

The LAD estimator minimizes absolute residuals,
\begin{equation}
    \beta_{i,t+1}^{\mathrm{LAD},W}
    =
    \arg\min_{\beta}
    \frac{1}{W}\sum_{s\in\mathcal W_{i,t}^{(W)}}
    \left|e_{i,s}^{(W)}(\beta)\right|,
    \label{eq:lad}
\end{equation}
consistent with mean-absolute-deviation estimation and hedging \cite{sharpe1971,liu2009mad}. The downside-semivariance estimator focuses on negative hedged outcomes,
\begin{equation}
    \beta_{i,t+1}^{\mathrm{Semi},W}
    =
    \arg\min_{\beta}
    \frac{1}{W}\sum_{s\in\mathcal W_{i,t}^{(W)}}
    \left[\max\{-e_{i,s}^{(W)}(\beta),0\}\right]^2,
    \label{eq:semi}
\end{equation}
following the lower-partial-moment and generalized-semivariance hedging literature \cite{lientse2000,chen2001gsv}.

The lower-tail quantile estimator uses pinball loss,
\begin{equation}
    \beta_{i,t+1}^{Q,W}
    =
    \arg\min_{\beta}
    \frac{1}{W}\sum_{s\in\mathcal W_{i,t}^{(W)}}
    \rho_{\tau_Q}\!\left(e_{i,s}^{(W)}(\beta)\right),
    \label{eq:quantile}
\end{equation}
where $\rho_{\tau}(u)=u(\tau-\mathbf 1\{u<0\})$ and $\tau_Q=0.10$ \cite{lien2016quantile}.

For Expected Shortfall, define hedging loss $\ell_{i,s}^{(W)}(\beta)=-e_{i,s}^{(W)}(\beta)$. The ES component uses the empirical CVaR/ES representation of Rockafellar and Uryasev \cite{rockafellar2000cvar},
\begin{equation}
\begin{aligned}
 \beta_{i,t+1}^{\mathrm{ES},W}
 &=\arg\min_{\beta}\Bigg\{q(\beta) +\frac{1}{(1-\alpha_{ES})W} \times
\\[-1pt]
 &\qquad  \sum_{s\in\mathcal W_{i,t}^{(W)}}
 [\ell_{i,s}^{(W)}(\beta)-q(\beta)]_+\Bigg\},
\end{aligned}
\label{eq:es}
\end{equation}
where $[x]_+=\max(x,0)$, $\alpha_{ES}=0.90$, and $q(\beta)$ is the corresponding empirical loss quantile. Minimum-CVaR hedge ratios have also been studied directly in futures hedging \cite{harris2006var}.

Collect the five candidate hedge ratios as
\begin{equation}
\bm{\beta}_{i,t+1}^{(W)}=\big[
 \beta_{i,t+1}^{\mathrm{MV},W},
 \beta_{i,t+1}^{\mathrm{LAD},W},
 \beta_{i,t+1}^{\mathrm{Semi},W},
 \beta_{i,t+1}^{Q,W},
 \beta_{i,t+1}^{\mathrm{ES},W}
\big].
\label{eq:components}
\end{equation}
\subsection{Training Targets and Candidate-Rank Prediction}

To learn candidate rankings, we use four observation-level loss criteria that can be evaluated directly from a single realized hedging residual,
\begin{equation}
\mathcal C=\{\mathrm{MSE},\mathrm{LAD},\mathrm{Semi},Q_{0.10}\}.
\label{eq:criterion_set}
\end{equation}
For candidate component $k$, let the one-step residual on target date $t+1$ be
\begin{equation}
    e_{i,t+1}^{(k,W)}
    =r_{i,t+1}-\beta_{i,t+1}^{(k,W)}r_{m,t+1}.
    \label{eq:oos_residual}
\end{equation}
The realized loss of candidate $k$ under criterion $c\in\mathcal C$ is
\begin{equation}
L_{i,t+1,c}^{(k,W)}=
\begin{cases}
(e_{i,t+1}^{(k,W)})^2, & c=\mathrm{MSE},\\[2pt]
|e_{i,t+1}^{(k,W)}|, & c=\mathrm{LAD},\\[2pt]
[\max\{-e_{i,t+1}^{(k,W)},0\}]^2, & c=\mathrm{Semi},\\[2pt]
\rho_{0.10}(e_{i,t+1}^{(k,W)}), & c=Q_{0.10}.
\end{cases}
\label{eq:criterion_losses}
\end{equation}
We use squared loss, absolute loss, semivariance loss, and quantile loss for criterion-specific learning because each can be evaluated for every realized hedging residual. Their sample averages are reported as MSE, LAD, downside semivariance loss, and Q10, respectively. Because squared loss and absolute loss preserve the same ordering of residual magnitudes, MSE and LAD produce the same within-stock-date rank target. We retain both criteria to provide one-to-one objective-matched comparisons with Rolling MV and Rolling LAD, respectively. Expected Shortfall instead summarizes losses over the tail of a distribution and therefore does not provide an equivalent single-observation loss for this ranking step. We retain ES as a component hedge estimator and an out-of-sample tail-risk measure, but not as a criterion-specific combination objective.

For each criterion, the five realized losses are ranked within each stock-date,
\begin{equation}
    R_{i,t+1,c}^{(k,W)}
    =\operatorname{Rank}_k\!\left(L_{i,t+1,c}^{(k,W)}\right),
    \qquad k=1,\ldots,5,
    \label{eq:ranks}
\end{equation}
where rank 1 is best. For each stock--date, criterion, and estimation window, the supervised target is therefore the five-dimensional rank vector
\begin{equation}
    \bm y_{i,t+1,c}^{(W)}
    =\bm R_{i,t+1,c}^{(W)}
    =\left[
    R_{i,t+1,c}^{(1,W)},\ldots,R_{i,t+1,c}^{(5,W)}
    \right]^{\top}.
    \label{eq:rank_target}
\end{equation}
This vector---not a realized beta---is the response variable used to train the ExtraTrees model. It is constructed only after the target-date returns are realized and is used only for model fitting and validation. At formation date $t$, the predictors contain no target-date return.

The predictors available at formation date $t$ include the five component hedge ratios, recent SPY volatility and drawdown, rolling stock--SPY correlation, cross-objective hedge-ratio dispersion, and CBOE VIX, VVIX, SKEW, and DSPX indexes observed by that date.

For each criterion $c$ and estimation window $W$, a multi-output ExtraTrees model maps the formation-date predictor vector to the five next-period rank predictions,
\begin{equation}
    \widehat{\bm R}_{i,t+1,c}^{(W)}
    =g_{\mathrm{ET}}^{(c,W)}(\bm z_{i,t}).
    \label{eq:extratrees}
\end{equation}
ExtraTrees is an ensemble of highly randomized decision trees that can capture nonlinearities and interactions without requiring feature scaling \cite{geurts2006extratrees}. We use 300 trees with maximum depth 10. The minimum leaf size is selected from $\{10,25,50\}$, and the softmax temperature used in the adaptive weighting rule is selected from $\{0.25,0.5,1,2\}$, using the 2021 validation period. The five entries of the rank vector are learned jointly in a multi-output regression. Thus, for a hedge formed at date $t$, the fitted model produces five predicted relative-performance scores for target date $t+1$ before $r_{i,t+1}$ and $r_{m,t+1}$ are observed. All selected hyperparameters are fixed before test evaluation.

\subsection{Adaptive, equal-weight, and hard-selection rules}

Predicted ranks are transformed into smooth convex weights,
\begin{equation}
    w_{i,t,c}^{(k,W)}
    =
    \frac{\exp\{-\widehat R_{i,t+1,c}^{(k,W)}/T\}}
    {\sum_{j=1}^{5}\exp\{-\widehat R_{i,t+1,c}^{(j,W)}/T\}},
    \label{eq:soft_weights}
\end{equation}
where $T$ controls weight concentration. As described above, $T$ is selected from $\{0.25,0.5,1,2\}$ using the 2021 validation period and is fixed before evaluation on the 2022--2024 test period. The adaptive hedge under criterion $c$ is
\begin{equation}
    \beta_{i,t+1,c}^{\mathrm{AW},W}
    =\sum_{k=1}^{5}w_{i,t,c}^{(k,W)}\beta_{i,t+1}^{(k,W)}.
    \label{eq:adaptive}
\end{equation}

Equal weighting uses the same five candidate ratios but no formation-date state variables or learned weighting rule,
\begin{equation}
    \beta_{i,t+1}^{\mathrm{EW},W}
    =\frac{1}{5}\sum_{k=1}^{5}\beta_{i,t+1}^{(k,W)},
    \label{eq:equal}
\end{equation}
and hard selection chooses the single candidate with the best predicted rank under criterion $c$,
\begin{equation}
    k_{i,t,c}^{*}=\arg\min_k\widehat R_{i,t+1,c}^{(k,W)},
    \qquad
    \beta_{i,t+1,c}^{\mathrm{Hard},W}=\beta_{i,t+1}^{(k_{i,t,c}^{*},W)}.
    \label{eq:hard}
\end{equation}
Equal weighting therefore tests whether combining the five candidate hedge ratios is useful without the broader predictor set used by the adaptive model. Hard selection tests whether choosing the single predicted-best candidate performs better than combining the candidates.

\section{Experimental Design}

\subsection{Data and chronological protocol}

The empirical application uses daily U.S. equity returns obtained from the Center for Research in Security Prices (CRSP) through Wharton Research Data Services (WRDS). The sample covers an S\&P 100 stock universe, with SPY used as the market hedge instrument. CBOE VIX, VVIX, SKEW, and DSPX indexes are obtained directly from Cboe Global Markets. Constituent and SPY returns are constructed consistently as daily total returns. Only observed paired returns are used, and missing returns are not forward-filled.

The experiment follows a chronological train--validation--test design. The objective-specific component hedge ratios do not require a separate training stage or hyperparameter tuning. At each formation date, they are computed directly from the most recent $W$ valid paired stock--SPY returns under their respective rolling objectives.

The train--validation--test split therefore applies to the ExtraTrees combination models. Target dates from 2010--2020 are used for training, 2021 is reserved for hyperparameter selection, and 2022--2024 forms the held-out out-of-sample test period. Split assignment is based on the target date, and all selected hyperparameters are fixed before testing. During the test period, the ExtraTrees model parameters remain fixed, while the rolling hedge ratios and predictor variables continue to update using information available at each formation date.

\subsection{Pre-specified estimation windows and rolling benchmarks}

The four estimation windows in Eq.~\eqref{eq:windows} are robustness settings rather than additional components. For a given $W$, all five candidate estimators use the same $W$ valid stock--SPY observations.

For each combination criterion, the proposed combinations are compared with a corresponding single-objective rolling hedge using the same estimation window:
\begin{enumerate}
    \item MSE: conventional Rolling MV (OLS);
    \item LAD: Rolling LAD;
    \item Downside: Rolling Downside/Semivariance; and
    \item Q10: Rolling 10\% quantile hedge.
\end{enumerate}
The MSE panel therefore uses the conventional minimum-variance OLS hedge as its rolling benchmark, while the remaining panels use the rolling estimator defined by the corresponding loss criterion. In every case, Obj. Gain is evaluated using the loss criterion of the panel. This construction keeps the estimation window and evaluation criterion fixed while testing whether combining hedge ratios estimated under different objectives improves on the corresponding rolling benchmark.

Within every objective--window block, the reported comparison methods are Adaptive-weight, Hard selection, Equal-weight, and the corresponding rolling estimator.

\subsection{Evaluation metrics}

The main variance-risk measure is stock-level hedge effectiveness,
\begin{equation}
    HE_i
    =1-
    \frac{\sum_{t\in\mathcal T_i}e_{i,t+1}^{2}}
    {\sum_{t\in\mathcal T_i}r_{i,t+1}^{2}},
    \label{eq:he}
\end{equation}
and the table reports the cross-sectional mean in percent. Higher HE is better. The 5\% Expected Shortfall, $ES_{5\%}$, is reported as a positive loss magnitude in percent, so lower values are better. Hedge-ratio turnover measures the average absolute change in the hedge ratio across valid stock--date observations:
\begin{equation}
    \operatorname{Turnover}
    =
    \frac{1}{|\mathcal O|}
    \sum_{(i,t)\in\mathcal O}
    \left|\beta_{i,t+1}-\beta_{i,t}\right|,
    \label{eq:turnover}
\end{equation}
where $\mathcal O$ denotes the set of valid stock--date observations in the out-of-sample test period. Lower turnover indicates a more stable hedge ratio over time. Turnover is reported in percent.

The objective-specific metric, Obj. Gain, is the percentage reduction in the panel loss relative to its corresponding rolling benchmark. Stock HE Win is the percentage of evaluated stocks with higher stock-level HE than the same benchmark. Positive Obj. Gain and higher Stock HE Win favor the combination method. Although MSE and LAD share the same ordinal rank target, their reported gains are evaluated using different losses and different rolling benchmarks.

Figure~\ref{fig:cumulative_abs} reports cumulative absolute hedging-loss saving for Adaptive-weight relative to the corresponding rolling benchmark. The same objective-matched comparison is used there: the MSE panel is relative to Rolling MV, while the LAD panel is relative to Rolling LAD. Each panel fixes the combination objective, and each line corresponds to one estimation window.
\section{Empirical Results}

\subsection{Objective-matched performance}

Table~\ref{tab:matched_objective_panels} reports benchmark-relative numerical results for the 60-, 90-, 120-, and 180-day windows. The clearest result is that Adaptive-weight produces positive Obj. Gain in every reported block. Under MSE, the gain is 0.27\%, 0.13\%, 0.10\%, and 0.08\% for 60, 90, 120, and 180 days. Under LAD, the corresponding gains are 0.22\%, 0.19\%, 0.16\%, and 0.21\%. These improvements are modest but consistently positive against the corresponding rolling benchmark.

The largest improvements occur for asymmetric downside criteria. Adaptive-weight Obj. Gain under downside semivariance is 2.01\%, 1.10\%, 0.75\%, and 0.50\% as the window increases from 60 to 180 days. Under Q10, the corresponding gains are 0.98\%, 0.38\%, 0.16\%, and 0.06\%. The stock-level HE comparisons are also broad: Adaptive-weight beats its corresponding rolling benchmark for 63.4--96.0\% of stocks across the sixteen reported objective--window blocks. The highest win rates occur for the downside and Q10 panels at the shorter windows.

The LAD, downside, and Q10 comparisons are objective matched. For example, the Q10 combination is compared with Rolling Q10 under the same pinball criterion, and the downside combination is compared with Rolling Downside under the same semivariance criterion. Positive Obj. Gain therefore indicates that a hedge constructed from estimates generated under several objectives can outperform the hedge estimated directly under the final evaluation objective.
\begin{table*}[t]
\centering
\scriptsize
\renewcommand{\arraystretch}{0.90}
\setlength{\tabcolsep}{2.15pt}
\caption{Out-of-sample hedging performance relative to corresponding rolling benchmarks.}
\label{tab:matched_objective_panels}
\begin{tabular}{rlrrrrr@{\hspace{1.25em}}rlrrrrr}
\toprule
\multicolumn{7}{c}{\textit{(a) Mean Squared Error}} & \multicolumn{7}{c}{\textit{(b) Least Absolute Deviation}} \\
\cmidrule(lr){1-7}\cmidrule(lr){8-14}
Window & Method & Gain & HE & ES$_{5\%}$ & Win & Turn. & Window & Method & Gain & HE & ES$_{5\%}$ & Win & Turn. \\
(day) & & (\%) $\uparrow$ & (\%) $\uparrow$ & (\%) $\downarrow$ & (\%) $\uparrow$ & (\%) $\downarrow$ & (day) & & (\%) $\uparrow$ & (\%) $\uparrow$ & (\%) $\downarrow$ & (\%) $\uparrow$ & (\%) $\downarrow$ \\
\midrule
60 & Adaptive-weight & \textbf{0.27} & \textbf{30.33} & \textbf{3.599} & \textbf{71.3} & \underline{2.75} & 60 & Adaptive-weight & \textbf{0.22} & \textbf{30.33} & \underline{3.599} & \textbf{64.4} & \textbf{2.75} \\
 & Hard selection & \underline{0.02} & \underline{30.11} & \underline{3.602} & \underline{55.4} & 4.20 &  & Hard selection & \underline{0.09} & \underline{30.11} & 3.602 & \underline{51.5} & 4.20 \\
 & Equal-weight & -0.12 & 30.08 & 3.613 & 46.5 & 2.85 &  & Equal-weight & -0.02 & 30.08 & 3.613 & 47.5 & \underline{2.85} \\
 & Rolling MV (OLS) & 0.00 & 30.06 & 3.603 & -- & \textbf{2.56} &  & Rolling LAD & 0.00 & 30.03 & \textbf{3.598} & -- & 3.12 \\
\cmidrule(lr){1-7}\cmidrule(lr){8-14}
90 & Adaptive-weight & \textbf{0.13} & \textbf{30.50} & \textbf{3.587} & \textbf{67.3} & \underline{1.84} & 90 & Adaptive-weight & \textbf{0.19} & \textbf{30.51} & \textbf{3.587} & \textbf{72.3} & \textbf{1.82} \\
 & Hard selection & -0.26 & 30.25 & \underline{3.590} & 46.5 & 3.33 &  & Hard selection & 0.01 & 30.29 & \underline{3.588} & 49.5 & 3.13 \\
 & Equal-weight & -0.07 & \underline{30.39} & 3.594 & \underline{53.5} & 1.85 &  & Equal-weight & \underline{0.07} & \underline{30.39} & 3.594 & \underline{59.4} & \underline{1.85} \\
 & Rolling MV (OLS) & \underline{0.00} & 30.30 & 3.591 & -- & \textbf{1.66} &  & Rolling LAD & 0.00 & 30.22 & 3.590 & -- & 2.04 \\
\cmidrule(lr){1-7}\cmidrule(lr){8-14}
120 & Adaptive-weight & \textbf{0.10} & \textbf{30.64} & \textbf{3.585} & \textbf{69.3} & 1.43 & 120 & Adaptive-weight & \textbf{0.16} & \textbf{30.64} & \textbf{3.585} & \textbf{68.3} & \underline{1.43} \\
 & Hard selection & -0.14 & 30.42 & 3.590 & 47.5 & 3.38 &  & Hard selection & -0.01 & 30.42 & 3.590 & 44.6 & 3.38 \\
 & Equal-weight & -0.05 & \underline{30.58} & 3.590 & \underline{61.4} & \underline{1.35} &  & Equal-weight & \underline{0.06} & \underline{30.58} & 3.590 & \underline{64.4} & \textbf{1.35} \\
 & Rolling MV (OLS) & \underline{0.00} & 30.44 & \underline{3.586} & -- & \textbf{1.21} &  & Rolling LAD & 0.00 & 30.35 & \underline{3.587} & -- & 1.47 \\
\cmidrule(lr){1-7}\cmidrule(lr){8-14}
180 & Adaptive-weight & \textbf{0.08} & \textbf{30.82} & \underline{3.582} & \textbf{72.3} & 1.04 & 180 & Adaptive-weight & \textbf{0.21} & \textbf{30.82} & \textbf{3.582} & \textbf{71.3} & 1.04 \\
 & Hard selection & -0.07 & 30.69 & 3.583 & 55.4 & 3.20 &  & Hard selection & 0.11 & 30.69 & \underline{3.583} & 52.5 & 3.20 \\
 & Equal-weight & -0.02 & \underline{30.77} & 3.586 & \underline{67.3} & \underline{0.92} &  & Equal-weight & \underline{0.14} & \underline{30.77} & 3.586 & \underline{65.3} & \textbf{0.92} \\
 & Rolling MV (OLS) & \underline{0.00} & 30.56 & \textbf{3.582} & -- & \textbf{0.82} &  & Rolling LAD & 0.00 & 30.47 & 3.585 & -- & \underline{1.02} \\
\midrule\addlinespace[1.5pt]
\multicolumn{7}{c}{\textit{(c) Downside Semivariance}} & \multicolumn{7}{c}{\textit{(d) 10\% Quantile Loss}} \\
\cmidrule(lr){1-7}\cmidrule(lr){8-14}
Window & Method & Gain & HE & ES$_{5\%}$ & Win & Turn. & Window & Method & Gain & HE & ES$_{5\%}$ & Win & Turn. \\
(day) & & (\%) $\uparrow$ & (\%) $\uparrow$ & (\%) $\downarrow$ & (\%) $\uparrow$ & (\%) $\downarrow$ & (day) & & (\%) $\uparrow$ & (\%) $\uparrow$ & (\%) $\downarrow$ & (\%) $\uparrow$ & (\%) $\downarrow$ \\
\midrule
60 & Adaptive-weight & \textbf{2.01} & \textbf{30.10} & \textbf{3.612} & \underline{96.0} & \underline{2.87} & 60 & Adaptive-weight & \textbf{0.98} & \textbf{30.34} & \textbf{3.599} & \textbf{95.0} & \textbf{2.75} \\
 & Hard selection & 1.79 & 29.65 & 3.615 & 73.3 & 6.44 &  & Hard selection & \underline{0.97} & \underline{30.15} & \underline{3.603} & 84.2 & 4.23 \\
 & Equal-weight & \underline{1.96} & \underline{30.08} & \underline{3.613} & \textbf{97.0} & \textbf{2.85} &  & Equal-weight & 0.60 & 30.08 & 3.613 & \underline{87.1} & \underline{2.85} \\
 & Rolling Downside & 0.00 & 29.09 & 3.651 & -- & 2.89 &  & Rolling Q10 & 0.00 & 29.32 & 3.649 & -- & 3.52 \\
\cmidrule(lr){1-7}\cmidrule(lr){8-14}
90 & Adaptive-weight & \textbf{1.10} & \textbf{30.40} & \textbf{3.593} & \underline{86.1} & 1.86 & 90 & Adaptive-weight & \textbf{0.38} & \textbf{30.50} & \textbf{3.587} & \textbf{83.2} & \textbf{1.85} \\
 & Hard selection & 0.95 & 30.07 & 3.598 & 57.4 & 5.60 &  & Hard selection & \underline{0.25} & 30.27 & \underline{3.591} & 66.3 & 3.42 \\
 & Equal-weight & \underline{1.09} & \underline{30.39} & \underline{3.594} & \textbf{87.1} & \underline{1.85} &  & Equal-weight & 0.21 & \underline{30.39} & 3.594 & \underline{81.2} & \underline{1.85} \\
 & Rolling Downside & 0.00 & 29.86 & 3.615 & -- & \textbf{1.81} &  & Rolling Q10 & 0.00 & 29.94 & 3.611 & -- & 2.29 \\
\cmidrule(lr){1-7}\cmidrule(lr){8-14}
120 & Adaptive-weight & \textbf{0.75} & \textbf{30.58} & \textbf{3.590} & \textbf{78.2} & 1.37 & 120 & Adaptive-weight & \textbf{0.16} & \textbf{30.62} & \textbf{3.587} & \textbf{78.2} & \underline{1.38} \\
 & Hard selection & 0.55 & 30.26 & 3.594 & \underline{47.5} & 4.92 &  & Hard selection & -0.01 & 30.45 & \underline{3.589} & 67.3 & 3.47 \\
 & Equal-weight & \underline{0.74} & \underline{30.58} & \underline{3.590} & \textbf{78.2} & \underline{1.35} &  & Equal-weight & \underline{0.09} & \underline{30.58} & 3.590 & \underline{74.3} & \textbf{1.35} \\
 & Rolling Downside & 0.00 & 30.26 & 3.604 & -- & \textbf{1.32} &  & Rolling Q10 & 0.00 & 30.21 & 3.609 & -- & 1.69 \\
\cmidrule(lr){1-7}\cmidrule(lr){8-14}
180 & Adaptive-weight & \textbf{0.50} & \textbf{30.77} & \textbf{3.586} & \underline{63.4} & 0.95 & 180 & Adaptive-weight & \textbf{0.06} & \textbf{30.77} & \textbf{3.586} & \textbf{70.3} & \underline{0.92} \\
 & Hard selection & 0.08 & 30.41 & 3.595 & 34.7 & 3.70 &  & Hard selection & -0.06 & 30.66 & \underline{3.586} & \underline{62.4} & 2.94 \\
 & Equal-weight & \underline{0.49} & \underline{30.77} & \underline{3.586} & \textbf{66.3} & \underline{0.92} &  & Equal-weight & \underline{0.05} & \underline{30.77} & 3.586 & \textbf{70.3} & \textbf{0.92} \\
 & Rolling Downside & 0.00 & 30.62 & 3.598 & -- & \textbf{0.90} &  & Rolling Q10 & 0.00 & 30.47 & 3.608 & -- & 1.15 \\
\bottomrule
\end{tabular}
\vspace{2pt}
\begin{minipage}{0.98\textwidth}
\scriptsize \textit{Notes:} Results are reported for 60-, 90-, 120-, and 180-day estimation windows. Gain is the percentage reduction in the panel loss and Win is the stock-level HE win rate, both relative to the corresponding rolling benchmark. For the MSE panel, the benchmark is the conventional rolling minimum-variance OLS hedge. Arrows indicate the preferred direction. Bold and underlined entries denote the best and second-best displayed values within each objective--window block.
\end{minipage}
\end{table*}

\subsection{Diversification versus adaptive weighting}

\begin{figure}[t]
    \centering
    \includegraphics[width=\columnwidth]
    {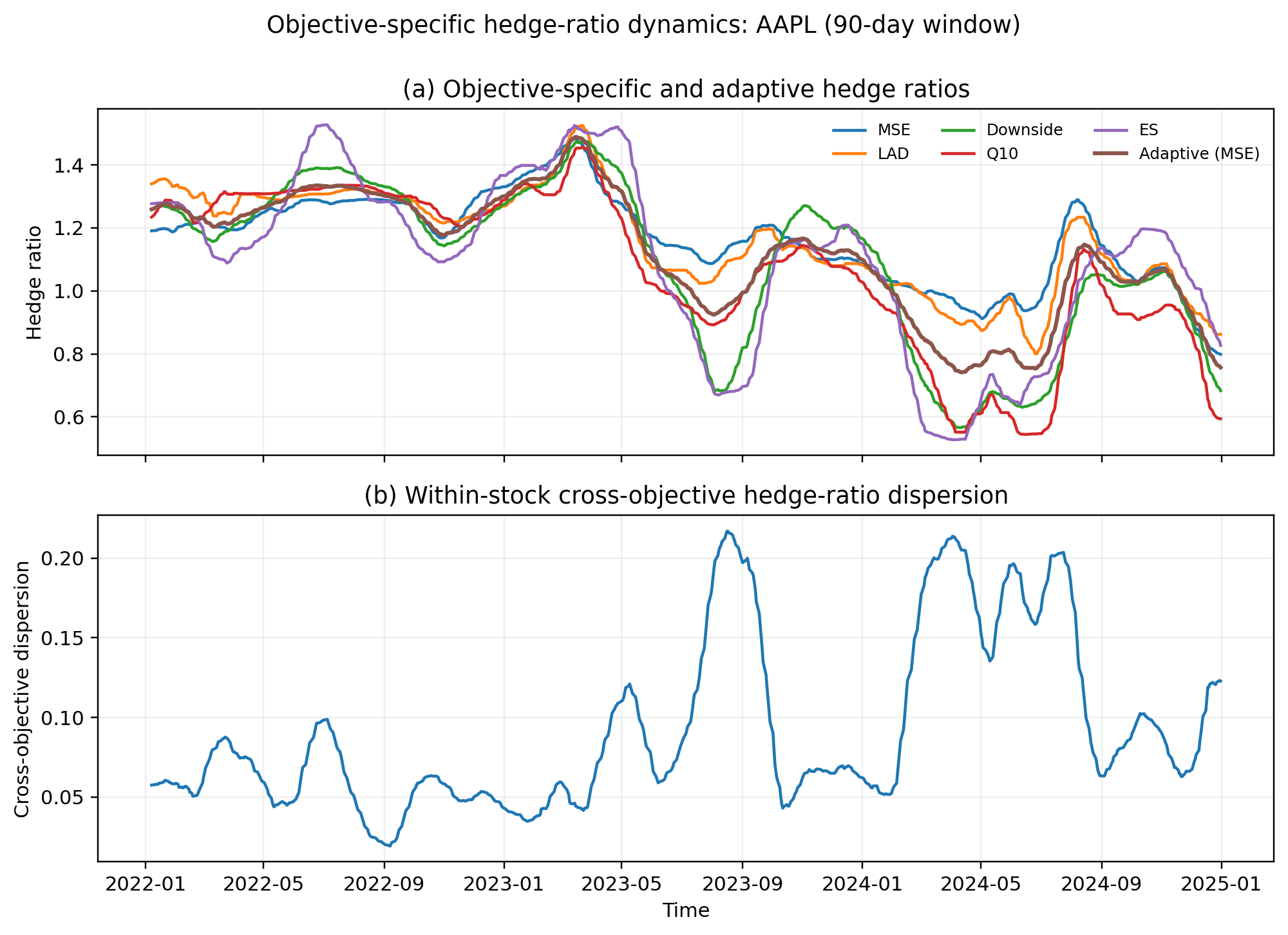}
    \caption{Objective-specific hedge-ratio dynamics for AAPL using the
    90-day estimation window. Panel (a) shows the objective-specific hedge
    ratios and the adaptive MSE hedge ratio. Panel (b) reports the
    corresponding cross-objective dispersion.}
    \label{fig:beta_dynamics}
\end{figure}

Figure~\ref{fig:beta_dynamics} illustrates how the five candidate hedge ratios evolve over time for AAPL using the 90-day estimation window. Panel~(a) shows that the five objective-specific rolling estimators, although fitted to the same stock--market return history, can generate materially different hedge-ratio paths. The MSE adaptive hedge remains a smooth combination of these interpretable candidates rather than an unrestricted direct hedge-ratio prediction. Panel~(b) reports the within-stock cross-objective dispersion, measured as the standard deviation across the five objective-specific hedge ratios at each formation date. The dispersion is strongly time varying, showing that differences among the five candidate hedge ratios also change over the out-of-sample period. AAPL is shown as an illustrative stock-level example; the cross-sectional evidence is reported in Table~\ref{tab:matched_objective_panels}.

Equal weighting provides a direct test of objective diversification without state-dependent learning. Under downside semivariance, Equal-weight delivers Obj. Gain of 1.96\%, 1.09\%, 0.74\%, and 0.49\% for 60, 90, 120, and 180 days. Under Q10, the corresponding gains are 0.60\%, 0.21\%, 0.09\%, and 0.05\%. Because Equal-weight uses only the objective-specific hedge ratios and no broader predictor set, these results show that part of the improvement is available from the estimates generated under different objectives themselves, rather than only from the adaptive learning model.

The same conclusion is weaker under symmetric objectives. Equal-weight MSE Obj. Gain is negative in all four reported windows, while LAD is slightly negative at 60 days and positive at 90, 120, and 180 days. Adaptive weighting improves on Equal-weight in every reported objective--window block, but the incremental advantage is often small when Equal-weight already performs strongly. Taken together, these comparisons distinguish the value available from a static combination of objective-specific candidates from the additional performance associated with state-dependent adaptive weighting.

Hard selection is generally less attractive than smooth weighting. Its turnover is materially higher in many blocks, especially under downside and quantile objectives, and its objective gains are usually below the Adaptive-weight values. Overall, smoothly combining the candidates performs better than switching entirely to the single predicted-best candidate.

\subsection{Risk metrics and objective dependence}

The HE and ES columns do not always move identically to Obj. Gain. For example, the 60-day LAD adaptive combination improves the LAD objective and stock-level HE, while its ES is marginally higher than that of Rolling LAD. By contrast, the downside and Q10 adaptive combinations reduce ES relative to their corresponding rolling benchmarks at the shorter windows. This is not contradictory: MSE, LAD, semivariance, quantile loss, stock-level HE, and ES emphasize different parts of the return distribution and aggregate observations differently. The benchmark-relative table is therefore more informative when read jointly rather than through a single headline metric.

\subsection{Cumulative absolute hedging-loss savings}
\begin{figure*}[t]
    \centering
    \includegraphics[width=0.8\textwidth]{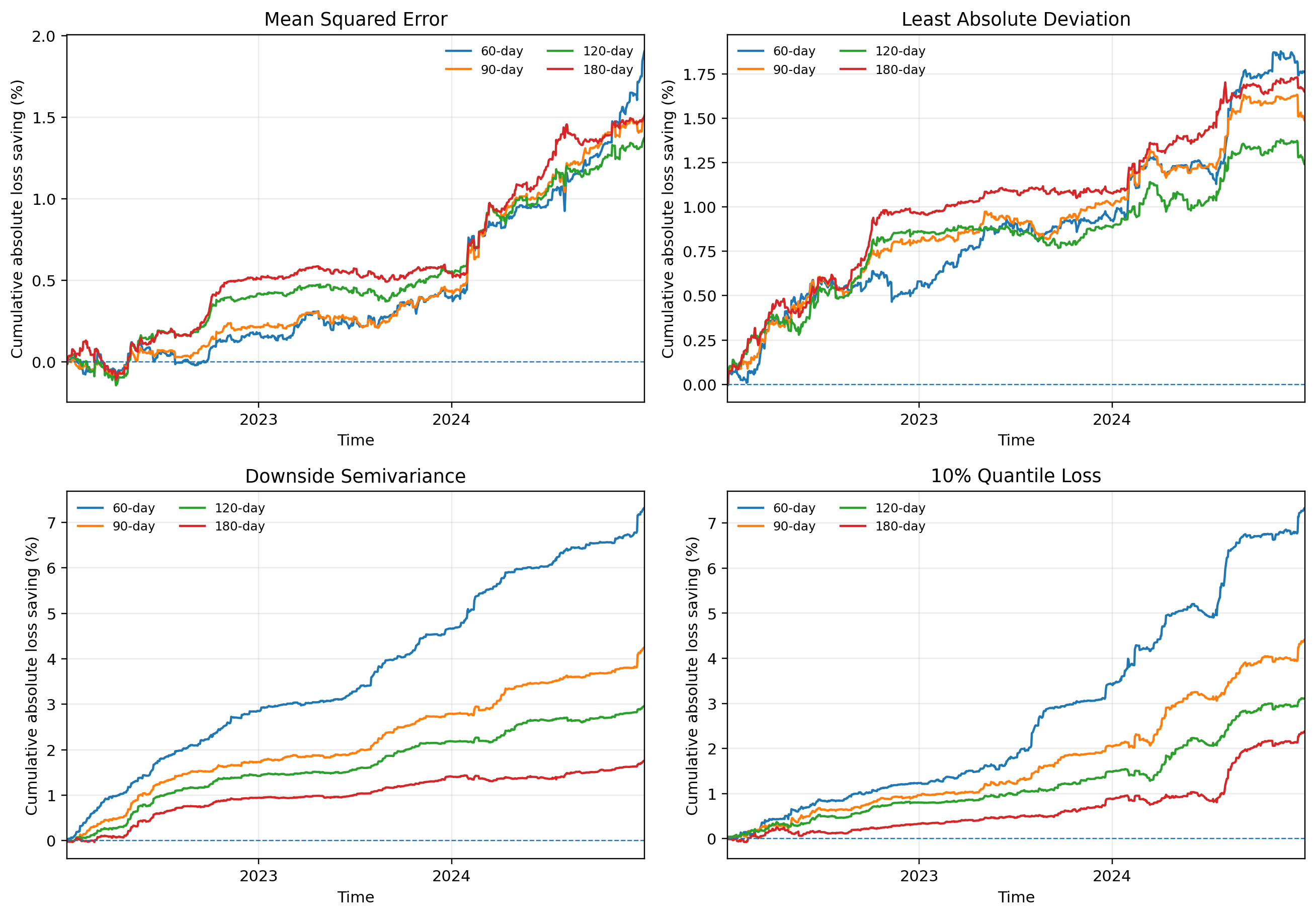}
    \caption{Cumulative absolute hedging-loss savings of Adaptive-weight relative to the corresponding rolling benchmark over the 2022--2024 out-of-sample test period. Each panel fixes the combination objective: Mean Squared Error, Least Absolute Deviation, Downside Semivariance, or 10\% Quantile Loss. Lines correspond to the pre-specified 60-, 90-, 120-, and 180-day estimation windows. Positive values favor Adaptive-weight.}
    \label{fig:cumulative_abs}
\end{figure*}

Figure~\ref{fig:cumulative_abs} reports cumulative absolute hedging-loss savings separately for each combination objective, with each line corresponding to one of the four estimation windows. All displayed paths finish above zero, indicating that Adaptive-weight accumulates positive absolute hedging-loss savings relative to the corresponding rolling benchmark by the end of the out-of-sample test period for every objective--window pair.

The effect of the estimation window differs across objectives. For downside semivariance and Q10, cumulative savings generally decrease as the window lengthens, with the 60-day window performing best and the 180-day window weakest. MSE and LAD show a less consistent ordering across windows. Thus, no single estimation window performs best across all objectives.

\section{Discussion}

The main finding is that choosing a final hedging objective does not necessarily imply that only the hedge ratio estimated directly under that objective is useful. This is clearest in the downside and Q10 panels, where Equal-weight improves the panel loss relative to the corresponding objective-matched rolling benchmark. Since Equal-weight uses only the five objective-specific hedge ratios and no broader formation-date state variables, these gains support the use of different objectives as a source of estimator diversity.

Second, adaptive weighting provides an additional improvement beyond the static combination. Adaptive-weight improves Obj. Gain over Equal-weight in every reported block, although the difference is small where Equal-weight is already strong. The adaptive model uses both the candidate hedge ratios and broader formation-date state information, so its incremental gain is best interpreted as the additional value of state-dependent combination.

Third, the effect of the estimation window varies across objectives. The cumulative-loss results show that shorter windows perform particularly well for downside semivariance and Q10, while the pattern is less pronounced for MSE and LAD. This suggests that asymmetric objectives may benefit more from recent information, whereas symmetric objectives are less sensitive to the choice of lookback window. Overall, the results indicate that the appropriate estimation horizon depends on the hedging objective rather than following a common pattern across all criteria.

Several limitations remain. The reported differences are descriptive out-of-sample comparisons, and we do not report statistical confidence intervals that account for serial and cross-sectional dependence. The design distinguishes static objective diversification from the full state-dependent combination, but it is not intended to identify the separate causal contribution of individual formation-date predictors. Expected Shortfall is retained as a candidate hedge estimator and evaluation metric but is not used as a criterion for candidate ranking. The four estimation windows are treated as pre-specified robustness settings rather than as a formal window-selection exercise.

\section{Conclusion}

This paper studies objective diversification of dynamic hedging. Five rolling estimators---minimum variance, LAD, downside semivariance, lower-tail quantile, and Expected Shortfall---map the same historical stock--market returns into different candidate hedge ratios. Separate ExtraTrees models then predict the next-period relative performance of these candidates under MSE, LAD, downside, and Q10 criteria and form smooth adaptive combinations.

The benchmark-relative results show that adaptive weighting improves the panel objective in every reported 60-, 90-, 120-, and 180-day block. The strongest gains occur for downside and Q10 criteria, where equal weighting---without the broader state predictors used by ExtraTrees---already captures much of the improvement, while MSE and LAD gains are smaller and depend more on adaptive reweighting. The cumulative-loss figure shows positive endpoint savings across all four objective panels and all four estimation windows, with the strongest short-window pattern under downside and Q10.

The broader implication is that choosing a hedging objective need not imply relying exclusively on the hedge ratio estimated under that objective. Different objectives can be used to generate alternative hedge-ratio estimates from the same return history, and these estimates can then be combined under the criterion relevant to the final hedging problem. In this sense, objective functions serve not only as optimization criteria but also as a source of estimator diversity.

\end{document}